\def\pictuph#1#2#3#4{%
\begin{figure}[htbp]
\begin{center}
\framebox[#1][l]{\raisebox{0cm}[0cm][#2]{ }}
\end{center}
\caption[]{\label{#3}\sloppy#4}
\end{figure}
}
\batchmode
\documentstyle[12pt]{article}
\begin{document}
\noindent
\hfill TTP93--17\\
\mbox{}
\hfill  June  1993   \\   
\vspace{0.5cm}
\begin{center}
  \begin{Large}
  \begin{bf}
  \vspace{3cm}
 RATES AND DISTRIBUTIONS IN TAU DECAY
 \footnote{\normalsize Supported by BMFT Contract 056KA93P}
\\
  \end{bf}
  \end{Large}

  \vspace{0.8cm}
  \begin{large}
   J.H. K\"uhn \\[5mm]
  \end{large}
    Institut f\"ur Theoretische Teilchenphysik\\
    Universit\"at Karlsruhe\\
    Kaiserstr. 12,    Postfach 6980\\[2mm]
    7500 Karlsruhe 1, Germany\\
  \vspace{4.5cm}
  {\bf Abstract}
\end{center}
%
\noindent
Semileptonic decays of polarised $\tau$ leptons are investigated.
Predictions for the rate, based on CVC and chiral Lagrangians, are
contrasted with experiments. Predictions for the angular distributions
of three meson final states are given. Emphasis is put on studies in
electron-positron annihilation where the neutrino escapes detection and
the $\tau$ restframe cannot be reconstructed. It is shown that the
form factors can be measured in ongoing high statistics experiments.
Of particular interest for the three meson case are the distribution
of the normal to the Dalitz plane and the distribution around this
normal. At LEP these distributions allow for an improved measurement of
the  $\tau$ polarisation. Implications are considered for an experiment
where the $\tau$
restframe is reconstructed. It is shown that the measurement of impact
parameters with the help of vertex detectors allows a full kinematic
reconstruction, including the direction of the $\tau$ and the missing
neutrino momentum.
\newpage
\section{Introduction}

Since the original discovery of the $\tau$-lepton
our understanding of its
production and decay properties has advanced
continuously through improved
experimental precision and through detailed
theoretical investigations.
All studies to date seem to be in excellent
agreement with the predictions
of the Standard Model (SM). Nevertheless there are a number
of reasons why precision studies
of the $\tau$-lepton ought to be pushed far
beyond the present level.
Pair production at $e^+e^-$-colliders
seems to be the most promising
experimental tool at present and in the
foreseeable future. The problems
and questions to be addressed in this connection
can be broken down into three
groups (Fig.\ref{ABC}).

\pictuph{16.9cm}{2.5cm}{ABC}%
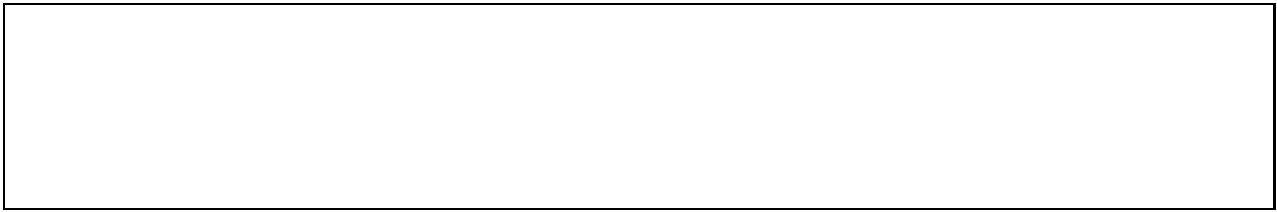
{{\em Left:} Production and decay of at $\tau$ pair in electron-positron
annihilation.\\ \mbox{\hspace{1.55cm}}
{\em Right:} CVC relating electron positron annihilation and $\tau$
decays.}
\noindent
{\bf A}
The {\em production mechanism}
through the virtual photon and the $Z_{0}$ boson
is unambigously fixed by the standard model.
The dependence
of the $\tau$ polarisation as measured at LEP
on the neutral vector coupling
constant can be exploited for a determination
of the weak mixing angle well
comparable in its precision to that of the
lepton forward backward asymmetry \cite{schmitt}.
Of course one may and should, in addition,
always test the SM, check lepton
universality (which could be invalidated in
certain variants of the two
Higgs doublet model \cite{Hoang}) or search
for an (CP violating) electric dipole moment \cite{wermes}.
These investigations are of course motivated
by the large $\tau$ mass which
might enhance effects that are suppressed by
the small masses in the case of
electron or muon.

\noindent
{\bf B}
Also the {\em charged current coupling}
of the $\tau$ to its neutrino and to the $W$
should be scrutinized as far as possible. The
determination of the neutrino
helicity, the measurement of the Michel
parameter or the search for a
small admixture of $V+A$ are obvious goals
in this context \cite {spaan}.

\noindent
{\bf C}
Semileptonic $\tau$ decays are a unique tool
for {\em low energy hadron physics}.
Amplitudes and rates of exclusive decay modes
can be predicted with chiral
Lagrangians, supplemented by information about
resonance parameters.
Conversely, these decay modes may develop into
a unique tool for the study
of $\rho,\rho'$ competing well with low energy
$e^{+}e^{-}$ colliders with
energies tuned to the region below 1.7~GeV.
$\tau$ decays allow, furthermore, the direct
production of resonances with
$J^{\mbox{\scriptsize PC}}=0^{-+}$ and $1^{++}$
with and  without strangeness,
a piece of information not directly accessible
from any  other experiment.\\

The topics {\bf A,B} and {\bf C}
are of course closely connected
 in any actual analysis.
Quite often theoretical or experimental
information on the decay has to be
included to pin down the production
amplitude or vice versa. Theoretical models
or low energy experiments (for example
ARGUS or CLEO) may therefore provide
important information for the $\tau$
analysis e.g. at LEP or for the analysis
of $\tau$'s from $W$ decay. Furthermore,
with increasing event rates, one may
study not only branching ratios or mass
distributions but one may also analyse
highly differential multidimensional
angular distributions leading to rigid
constraints on the theoretical input in the analysis.

These themes will be illustrated in the
following sections by a few typical
examples. In section 2 predictions
for the rate and for the spectral
functions will be discussed. The important
role of angular distributions will
be emphasized in section 3 and the
important notion of structure
functions will be introduced with emphasis
on investigations based on momenta
of the final hadrons only. The possibilities
for full reconstruction of the
kinematics, despite the missing neutrinos with
the help of impact parameter
measurements will be demonstrated in section 4.
\section{Decay Rates and hadron physics}
{\em Semileptonic rates and CVC}\\
The production of $1^{--}$ states with
strangeness zero in $\tau$ decay and in
$e^{+}e^{-}$ annihilation are intimitely
connected (Fig.\ref{ABC}). Semileptonic
$\tau$ decays explore the mass region
between $2m_\pi$ and $m_\tau$
with systematic errors different form
those of $e^+e^-$ experiments.
Already now they allow for example, a
determination of the pion form factor
with an accuracy comparable to a large
number of low energy $e^{+}e^{-}$
experiments. To wit, the experimental
results for the ratio
$\mbox{Br}(\tau^{-}\rightarrow \pi^{-}\pi^{0}\nu)/\mbox{Br}
(\tau^{-}\rightarrow e\overline{\nu}\nu)$
of ARGUS \cite{ARGUS} ($1.31 \pm 0.06$)
and ALEPH \cite{ALEPH} ($1.38 \pm 0.04$) are in nice
agreement with the CVC prediction
\cite{santa} of $1.32\pm 0.05$. The impact of
an improved pion form factor determination
(and the corresponding information on $4\pi$,
$\omega\pi$,...) for $g-2$
predictions of the muon and for the running of
$\alpha_{\mbox{\scriptsize QED}}$ are evident.

Based on the relation between the pion form
factor, the
$\tau$ decay rate and $\pi^{+}\pi^{-}$
 production in $e^{+}e^{-}$ collisions,
one may deduce a relation between the
differential effective luminosity from
$\tau$ decays and the true luminosity of a ``$\tau$ factory''
${\rm L}_\tau$.
\newcommand\qm{\frac{Q^2}{m_\tau^2}}
\begin{equation}\label{dl}
\frac{1}{{\rm L}_\tau}
\frac{d{\rm L}_{\mbox{\scriptsize eff}}}
{d(Q^{2}/m_{\tau}^{2})} \;=\; \frac{\Gamma_e}{\Gamma_\tau}
\cos\theta_c\qm(1-\qm)^2(1+2\qm)\frac{4m_\tau^2}{s_\tau}
(1+2\frac{m_\tau^2}{ s})\beta_\tau
\end{equation}
\pictuph{16.9cm}{6.5cm}{lumi}%
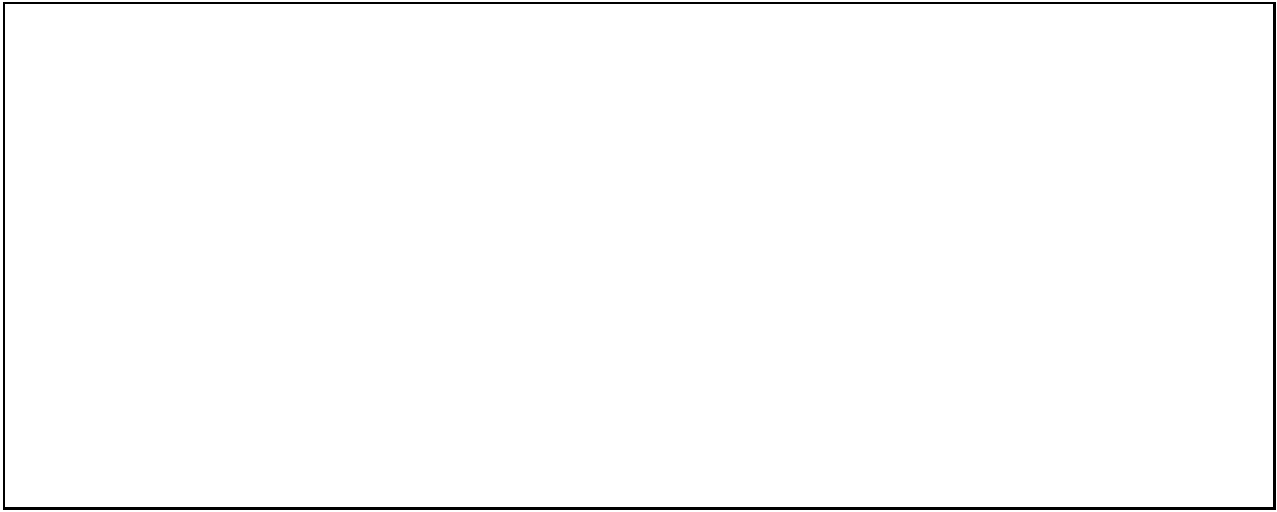
{{\em Left:} Differential effective luminosity at 4.2 GeV as defined in
(\ref{dl}).\\ \mbox{\hspace{1.55cm}}
{\em Right:} Normalised structure functions $w_C/w_A$, $w_D/w_A$ and
$w_E/w_A$ as functions of $Q^2$.}
On top of the $Z$ resonance
this result has to be multiplied
by $9\times\mbox{Br}(Z\to\tau^+\tau^-)^2/\alpha^2$.
The differential luminosity is shown in
Fig. \ref{lumi} for $\sqrt{s}=4.2 {\rm GeV}$.
The shape of this curve is of course
energy independent, the scale is modified
by a factor 0.24 for $\sqrt{s}=10{\rm GeV}$
and by 0.56 on top of the $Z$ resonance, respectively.\\
\noindent
{\em Axial current}\\
The theoretical prediction for
three pion and quite generally for three meson
states is on less safe grounds,
and the rates cannot be related to other
experimental observables. For small
pion momenta one may invoke the prediction
based on the chiral Lagrangian, which
determines the form and the normalisation of the
relevant matrix element \cite{FWW}
\begin{equation}
\langle\pi^-(p_1)\pi^-(p2)\pi^+(p_3)|{\cal J}_\alpha(0)|0\rangle
\equiv J_\alpha=
-i\frac{2\sqrt{2}}{3 f_\pi}(g_{\alpha\beta} - \frac{Q_\alpha
Q_\beta}{Q^2})
((p_1^\alpha - p_3^\alpha) + (p_2^\beta - p_3^\beta) )
\end{equation}
This parameterfree prediction is reliable
in the region of small
$Q^2=(p_1 + p_2 + p_3)^2$ where one is,
however, confronted with low
statistics. For increasing $Q^2$ and
increasing $s_k=(p_i + p_j)^2$
these calculations have to be supplemented
by form factors deduced from vector
dominance models.
This introduces some dependence on factors
which can be studied and explored
experimentally by investigating distributions
in the Dalitz plot and angular
distributions. Satisfactory agreement between
theory and experiment is obtained
for the three pion mode \cite{santa}. Predictions
for three meson states involving
kaons and eta's are also available \cite{decker}.
The rates and shapes of the
distributions exhibit a more drastic dependence
on the presently unknown form
factors.

The decay rates involving three different
mesons (for example $K^{-}\pi^{-}K^{+}$,
$K^{-}\pi^{-}\pi^{+}$ or $\eta \pi^{-} \pi^0$)
allow for axial vector and vector current
induced amplitudes at the same time.
The latter can be related to the anomaly
\cite{Kramer,decker}, the former is again
given by the chiral Lagrangian. As a
consequence of the large kaon mass and
the high threshold the predictions are
fairly sensitive towards the (model
dependent) assumptions on the interpolating
form factors. These form factors
have been implemented in the Monte Carlo
program TAUOLA \cite{was} which allows to test
and simulate the distributions in the Dalitz
plot as well as the angular distributions discussed below.

\section{Structure functions and angular distributions}

The information that can be deduced from
a full analysis is encoded in the hadronic tensor
$H_{\mu\nu}=J_\mu J^*_\nu$.
For a final state consisting of (pseudo-)%
scalar mesons only, $H_{\mu\nu}$ corresponds
to 16 real functions of $Q^2$, $s_1$ and $s_2$.
For a three pion state in
the spin one configuration, the most general current
is given by
\begin{equation}
J_\alpha=
(g_{\alpha\beta} - \frac{Q_\alpha Q_\beta}{Q^2})
((p_1^\alpha - p_3^\alpha)F(s_1,s_2,Q^2) + (1\longleftrightarrow2) )
\end{equation}
In the three pion rest frame $J_{0}=0$
and $\vec{J}$ is confined to the
plane spanned by $\vec{q}_{1}$ and
$\vec{q}_{2}$ and therefore has only two
independent components. The tensor
$H_{\mu\nu}$ is therefore determined by four
real functions, which in turn allow
to reconstruct the form factors $F$
introduced above. In the hadronic rest frame and with the
coordinates 1 and 2 in the $\vec q_1$,$\vec q_2$ plane
a convenient choice
for the structure functions $W_{i}$ which build up the hadronic tensor
reads as follows
\begin{eqnarray*}
W_{A} &=& H^{11} + H^{22} \\
W_{C} &=& H^{11} - H^{22} \\
W_{D} &=& 2\mbox{Re} H^{12} \\
W_{E} &=& 2\mbox{Im} H^{12} \\
\vspace{-1cm}
\end{eqnarray*}
$W_{A}$ governs the rate and the
distribution in the Dalitz plot.
The angular
distributions are most easily characterized
in the hadron rest frame.
$W_{C}$ and $W_{E}$ multiply the odd and
even parts of the distribution of
the normal on the three pion plane
and $W_{D}$ describes the rotations within
this plane. (In passing it should be
mentioned that also $\tau \rightarrow
\nu \pi \omega(\rightarrow 3\pi)$ can
be studied with similar techniques.)
For an unpolarized $\tau$ one predicts
\begin{equation}\label{distr}
\newcommand\mq{\frac{m_\tau^2}{Q^2}}
\frac{dN}{d\cos\beta d\gamma}\propto
[(1-\mq)(1+\cos^2\beta) + 2\mq] W_A
-(1-\mq)\sin^2\beta (\cos 2\gamma W_C - \sin 2\gamma W_D)
+ 2 \cos\beta W_E
\end{equation}
with the angles $\beta$ and $\gamma$
as defined in Fig.\ \ref{euler}.
\pictuph{16.9cm}{7cm}{euler}%
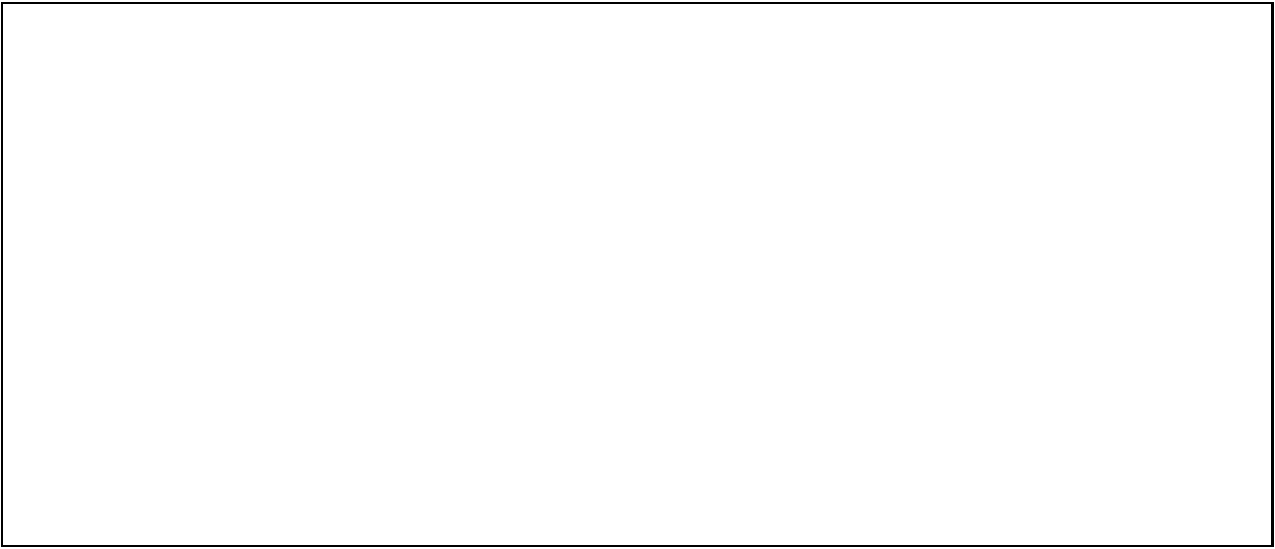
{{\em Left:} Definition of the angles $\beta$ and $\gamma$ if
the rest frame of the $\tau$ is known. \\ \mbox{\hspace{1.55cm}}
{\em Right:} Nonrelativistic
illustration $\vec s$ and of the angles $\psi$, $\theta$.}
The structure functions, averaged over
$s_1$ and $s_2$, are shown in Fig.\ \ref{lumi} as functions of
$Q^2$.

In practice the $\tau$ rest frame
and hence $\vec{n}_{\tau}$, that is the direction
of the $\tau$ as seen from the hadron
rest frame, are unknown as a consequence
of the missing neutrino momentum. However,
one may in this case replace
$\vec{n}_{\tau}$ by $\vec{n}_{L}$,
the direction of the lab as seen from the
hadron rest frame.The analysis then
involves the angle $\theta$ between
$\vec{n}_{\tau}$(lab), (the direction
of flight of the $\tau$ as seen from the
lab), and $\vec{n}_{H}(\tau)$
(the direction of the hadronic system as seen
from the $\tau$) and, furthermore,
the angle $\psi$ between the lab and the $\tau$
directions, as seen from the
hadronic system. Both angles can be expressed by
$x$, the energy of the hadronic system in the lab
\begin{eqnarray*}
\cos\theta & = & \vec{n}_{\tau}(\mbox{lab})\vec{n}_{H}(\tau) \;=
\frac{2x m^2_\tau -m^2_\tau - Q^2}
{(m^2_\tau-Q^2)\sqrt{1-4m^2_\tau/s}}\\
\cos\psi & = & \vec{n}_{L}(\mbox{had})\vec{n}_{\tau}(\mbox{had}) \;=
\frac{x( m^2_\tau + Q^2)  - 2 Q^2}
{(m^2_\tau-Q^2)\sqrt{1-4m^2_\tau/s}}
\end{eqnarray*}
The importance of the angle $\theta$ was originally observed
in \cite{tsai}, the angle
$\psi$ was introduced in \cite{KW}.
In the nonrelativistic approximation
the relation between $\theta,\psi$ and
$\theta_{L}$ is indicated in Fig.\ \ref{euler}.
Including longitudinal $\tau$
polarisation $P_\tau$ the experimentally
observable angular distribution can be cast into the following form
\cite{KM1,KM2}:
\begin{eqnarray*}
\frac{dN}{d\cos\theta d\cos\beta d\gamma}&\propto&
[K_1(1+\cos^2\beta) + 2K_2 - (K_1\sin^2\psi + K_4\sin2\psi)
(3\cos^2\beta - 1)/2] W_A\\
&&-[K_1(3\cos^2\psi - 1)/2 - 3/2 K_4\sin2\psi]
\sin^2\beta (\cos^2\gamma W_C - \sin^2\gamma W_D)\\
&&+ 2 [K_3\cos\psi - K_5\sin\psi] \cos\beta W_E
\end{eqnarray*}
with the coefficients $K_{i}$ defined as functions of
$\theta$, $\psi$ and of $P_\tau$.
The most general case, including a
spin zero contribution of the hadronic
matrix element, and the
corresponding predictions relevant for
the general three mesons
final state can be found in \cite{KM2}.
These distributions can be exploited
to determine the $\tau$ polarisation in
$Z$ decays.
Experimental studies of the sensitivity
of the method demonstrate that the
sensitivity is increased  from about
0.23 to 0.45 (corresponding to a
statistical error of $1/0.45\sqrt{N_{evt}}$)
if the full angular distribution
is incorporated \cite{davier,privt,aleph2}. The dependence on
the hadronic matrix element is weak
and could furthermore be reduced by the
corresponding measurements at ARGUS
and CLEO.
\\

\section{Tau kinematics from impact parameters}
\newcommand\beq{\begin{equation}}
\newcommand\eeq{\end{equation}}
\newcommand\tlp{{\theta^L_+}}
\newcommand\tlm{{\theta^L_-}}
\newcommand\cp{\cos\tlp}
\newcommand\cm{\cos\tlm}
\newcommand\spp{\sin\tlp}
\newcommand\sm{\sin\tlm}
\newcommand\np{\vec n_+}
\newcommand\nm{\vec n_-}
\newcommand\Pp{\vec p_+}
\newcommand\Pm{\vec p_-}
\newcommand\cph{\cos\varphi}
\newcommand\sph{\sin\varphi}
\newcommand\vd{\vec d}
\newcommand\dmin{\vd_{min}}
%
\pictuph{14cm}{5cm}{Kin}%
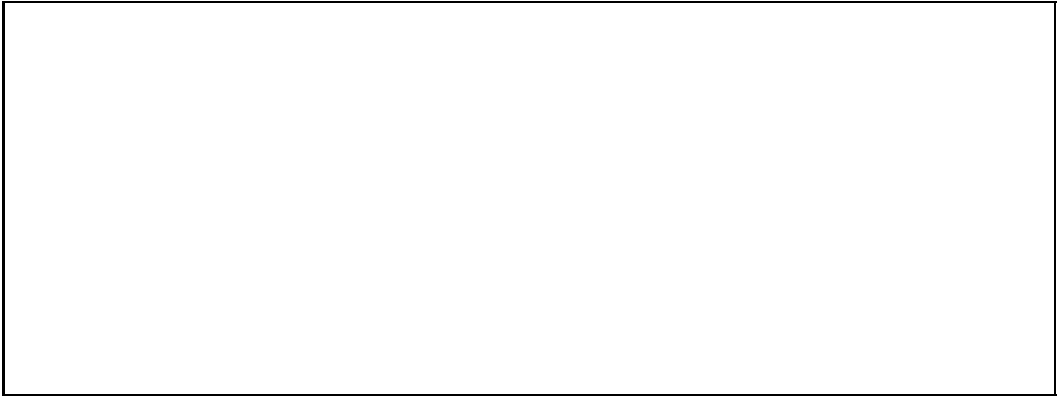
{Kinematic configuration indicating the relative orientation of the
hadronic tracks, the $\tau$ directions and the vector $\dmin$.}

A further increase in sensitivity could
be achieved if the $\tau$ direction
could be reconstructed experimentally,
such that $\vec{n}_{L}$ could be replaced by
$\vec{n}_{\tau}$ and the simpler eq.\ \ref{distr} would be
applicable. Theoretical predictions of
the full hadron distribution from the
decay of an arbitrarily polarised
$\tau$ are given in appendix B of \cite{KM2}.

As shown in \cite{kimp} the $\tau$ direction
can be reconstructed if the hadron tracks
are measured with the help of microvertex
detectors, even if the production
vertex is unknown as a consequence of the
large beam spot.

Let us assume that
both $\tau$ decay into one charged hadron each and that both
charged tracks can be measured with high precision.  The
direction $\vec d_{min}$
of the minimal distance between the two nonintersecting
charged tracks (Fig.\ref{Kin}) resolves the ambiguity and introduces
two additional constraints that can be used to reduce the measurement
errors.  The $\tau^+$ and $\tau^-$ decay points and their original
direction of flight can then be determined as follows.

The angles $\theta^L_{\pm}$ between the
$\tau^\pm$ and the  hadron $h^\pm$
directions respectively as defined in the lab frame are
given by the energies of $h^+$ and $h^-$ \cite{KW}:
\beq
\cm=\frac{\gamma x_- - (1+r^2_-)/2\gamma}
                      {\beta\sqrt{\gamma^2x_-^2-r_-^2}}
\eeq
\beq
\sm=
    \sqrt{\frac{(1-r_-^2)^2/4 - (x_--(1+r^2_-)/2)^2/\beta^2}
                      {\gamma^2x_-^2-r_-^2}}
\eeq
\beq
x_-=E_{h^-}/E_\tau \hspace{2cm}   r_-=m_{h^-}/m_\tau
\eeq
and similarly for $\cp$ and $\spp$.

The velocity
$\beta$, and the boost factor $\gamma$ refer to the $\tau$ in the lab
frame.

The original $\tau^-$ direction must therefore lie on the cone of
opening angle $\tlm$ around the direction of $h^-$ and on the cone of
opening angle $\tlp$ around the reflected direction of $h^+$.  The
extremal configuration
where $\tlp$ or $\tlm$ assume the values 0 or $\pi$,
or where the
two cones touch in one line, leads to a unique solution for the $\tau$
direction.  In general a twofold ambiguity arises, as
is obvious from this geometric argument. The
cosine of the relative azimuthal angle $\varphi$ between the directions
of $h^+$ and $h^-$ denoted by $\np$
and $\nm$ can be calculated from the momenta
and energies of $h^+$ and $h^-$ as follows:  In the coordinate frame
(see Fig.\ref{Kin}) with the $z$ axis
pointing along the direction of $\tau^-$
and with $\nm$ in the $xz$ plane and positive $x$ component
\beq
\frac{\Pm}{|\Pm|}\equiv\nm=
\left(\begin{array}{c} \sm    \\0       \\\cm     \end{array}\right)
\hspace{1cm}
\frac{\Pp}{|\Pp|}\equiv\np=
\left(\begin{array}{c}\spp\cph \\ \spp\sph\\-\cp    \end{array}\right)
\eeq
and $\cph$ can be determined from
\beq
\nm\np=-\cm \cp + \sm \spp\cph
\eeq
The well-known twofold ambiguity in $\varphi$ is evident from this formula.

Additional information can be drawn from the precise
determination of tracks close to the production point.
Three-prong decays allow to reconstruct the decay vertex and the
ambiguity can be trivially resolved.

However, single-prong events may also serve this purpose.  Let us
first consider decays into one charged hadron on each side.
Their tracks and in particular the vector $\dmin$ of closest
approach
(Fig.\ref{Kin}) can be measured with the help of microvertex
detectors. The vector pointing
from the $\tau_-$  to the $\tau_+$ decay vertex
\beq
\vec d\equiv \vec \tau_+ -\vec\tau_-= -l
\left(\begin{array}{c}0\\0\\1\end{array}\right)
\eeq
is oriented  by definition into the negative $z$ direction
($l>0)$.
The vector $\dmin$ can on the one hand be measured, on the other hand
calculated from $\vec d$, $\np$ and $\nm$:
\beq\label{eqdmin}
\dmin=\vd\, + \,
[(\vd\np\,\np\nm - \vd\nm)\nm + (\vd\nm\,\np\nm - \vd\np)\np]
/(1-(\nm\np)^2)
\eeq
The sign of the projection of $\dmin$ on $\np\times\nm$ then
determines the sign of $\varphi$ and hence resolves the
ambiguity.
\beq
\dmin(\np\times\nm) = l \spp\sm\sph
\eeq
The length of the projection determines $l$ and hence
provides a measurement of the lifetimes
of
$\tau_+$ plus $\tau_-$.
Exploiting the fact that $\vd\nm=-l\cm$ and $\vd\np=l\cp$ the direction
of $\vd$ can be geometrically constructed by inverting (\ref{eqdmin}):
\beq
\vd/l=\dmin/l \, - \,
[(\cp \, \np\nm + \cm)\nm + (-\cm \, \np\nm - \cp)\np] / (1-(\nm\np)^2)
\eeq

The generalization of this method to decays into multihadron
states with one or several neutrals
and a more detailed discussion of the constraints resulting from
this method can be found in \cite{kimp}.
\sloppy
\raggedright

\end{document}